\documentclass[prb,twocolumn]{revtex4}
\usepackage{amssymb}
\usepackage{graphicx}
\usepackage{amsfonts}
\usepackage{amsmath}

\begin{document}

\title{Adiabatic quantum pumping and rectification effects in interacting quantum dots}

\author{F. Romeo$^1$ and R. Citro$^{1,2}$}
\affiliation{$^1$Dipartimento di Fisica ''E. R. Caianiello'' and
C.N.I.S.M., Universit{\`a} degli Studi di Salerno, Via S. Allende,
I-84081 Baronissi (Sa), Italy\\
$^2$Labratorio Regionale SuperMat, CNR-INFM, Via S. Allende,
I-84081 Baronissi (Sa), Italy}

\begin{abstract}
We derive a formula describing the adiabatically pumped charge
through an interacting quantum dot within the scattering matrix
and Green's function approach. We show that when the tunneling
rates between the leads and the dot are varied adiabatically in
time, both in modulus and phase, the current induced in the dot
consists of two terms, the pumping current and a
rectification-like term. The last contribution arises from the
time-derivative of the tunneling phase and can have even or odd parity
with respect to the pumping phase $\varphi$. The
rectification-like term is also discussed in relation to some
recent experiments in quantum-dots.
\end{abstract}

\pacs{73.23.-b,72.10.Bg}

\keywords{adiabatic pumping, quantum dot, rectification effect}

 \maketitle

\section{ Introduction}
The idea of quantum pumping, i.e. of producing a dc current at
zero bias voltage by time periodic modulation of two system
parameters, dates back to the work of Thouless \cite{thouless}. If
the parameters change slowly as compared to all internal time
scales of the system, the pumping is {\it adiabatic}, and the
average charge per period does not depend on the detailed time
dependence of the parameters. Using the concept of emissivity
proposed by B\"{u}ttiker et al.\cite{buttiker_injectivity},
Brouwer\cite{brouwer98} related the charge pumped in a period to
the derivatives of the instantaneous scattering matrix of the
conductor with respect to the time-varying parameters. Since then,
a general framework to compute the pumped charge through a
conductor has been developed for noninteracting
electrons\cite{various}. The interest in the pumping phenomenon
has shifted then to the experimental\cite{exp-dot} investigations
of confined nanostructures, as quantum dots, where the realization
of the periodic time-dependent potential can be achieved by
modulating gate voltages applied to the structure\cite{exp-dot}.
In case of interacting electrons the computation of the pumped
charge becomes rather involved and few works have addressed this
issue for different systems\cite{various_interacting} and in
specific regimes. As for the case of interacting quantum dots, the
pumped charge in a period was calculated by Aono\cite{aono_qd} by
exploiting the zero-temperature mapping of the Kondo problem. A
very general formalism was developed in
Ref.[\onlinecite{fazio_prl_05}] where an adiabatic expansion of
the self-energy based on the average-time approximation was used
to calculate the dot Green's function while a linear response
scheme was employed in Ref.[\onlinecite{sela_prl_06}]. More
recently, another interesting study\cite{silva_07} was performed
aiming at generalizing Brouwer's formula for interacting systems
to include inelastic scattering events.

In this work we present a general expression for the adiabatic
pumping current in the interacting quantum dot in terms of
instantaneous properties of the system at equilibrium,
generalizing the scattering approach for noninteracting particles
and discuss the limit of its validity. To get a pumped current the
two model parameters which are varied in time  are the tunneling
rates between the noninteracting leads and the quantum dot. In
particular, we let them vary both in modulus and phase through the
adiabatic and periodic modulation of two external parameters (e.g.
gate voltages or magnetic fields) and show that a
rectification-like term arises in the current due to the
time-dependent tunneling phase.

The plan of the paper is the following. In Sec.\ref{sec:model} we
introduce the model and relevant parameters. We develop the
scattering matrix approach together with the Green's function
formalism to derive the formula of the pumped current through an
interacting multilevel quantum dot in the adiabatic regime at very
low-temperatures. In Sec.\ref{sec:single-level} we specialize on a
single-level quantum dot and give the explicit expression of the
pumped current. Conclusions are given in Sec.\ref{sec:conclusions}.

\section{ The model and formalism}
\label{sec:model}

 We consider a multi-level quantum dot (QD)
coupled to two noninteracting leads, with the external leads being
in thermal equilibrium. The Hamiltonian of the system is given by:
\begin{equation}
\label{eq:ham} H=H_{leads}+H_{dot}+H_{tun},
\end{equation}
where $H_{leads}=\sum_{k,\sigma,\beta}\epsilon_{\beta}(k)
c^\dagger_{k,\sigma,\beta}c_{k,\sigma,\beta}$, with
$c^\dagger_{k,\sigma,\beta}(c_{k,\sigma,\beta})$ the creation
(annihilation) operator of an electron with spin
$\sigma=\uparrow,\downarrow$ in the lead $\beta=L,R$ and
dispersion $\epsilon_{\beta}(k)$. The QD is described by the
Hamiltonian $H_{dot}=\sum_{j\sigma} \epsilon_{j\sigma} n_{j\sigma}
+U n_{j\uparrow} n_{j\downarrow}$, where
$n_{j\sigma}=d^\dagger_{j\sigma} d_{j\sigma}$ with
$d^\dagger_{j\sigma} (d_{j\sigma})$ the creation (annihilation)
operator of the electron with spin $\sigma$ and
$\epsilon_{j\sigma}$ the dot $j$-th energy level. The on-site
energy $U$ describes the Coulomb interaction. The tunneling
Hamiltonian is given by $H_{tun}=\sum_{k,\sigma,\beta,j}\lbrack
V_{k,\sigma,\beta,j}(t) c^\dagger_{k,\sigma,\beta}d_{j\sigma}+
H.c. \rbrack$, with time-dependent tunnel matrix elements
$V_{k,\sigma,\beta,j}(t)$. For simplicity we assume that
$V_{k,\sigma,\beta,j}$ are spin independent, i.e.
$V_{k,\sigma,\beta,j}=V_{\beta,j}$ and that both the modulus and
the phase of $V_{\beta,j}(t)$ vary in time with frequency
$\omega$, i.e.
$V_{\beta,j}(t)=|V_{\beta,j}(t)|\exp\{i\Phi_{\beta}(t)\}$. Their
explicit time dependence is determined by two external parameters
(e.g. two gate voltages applied at the barriers of the dot or a
gate voltage and a magnetic field) which are varied adiabatically
and periodically in time or by the presence of parasitic bias
voltages. Two specific examples will be considered below. In
particular, we will specialize on the case in which the tunneling
phase can vary harmonically or linearly in time.
 The instantaneous
strength of the coupling to the leads is instead characterized by
the parameters $\Gamma^\beta_{nm}(t,t)=2\pi\rho V_{\beta,m}(t)
V^{\ast}_{\beta,n}(t)$, where $\rho$ is the density
of states in the leads at the Fermi level.  
By varying in time $V_{L,j}(t)$ and $V_{R,j}(t)$ and keeping them
out of phase, the charge $\mathcal{Q}$ pumped in a period $T$ is
related to the time dependent current $I^L(\tau)$ flowing through
the left barrier, i.e. $\mathcal{Q}=\int_0^T d \tau I^L(\tau)$.

While the exact formula for the current depends on time-dependent
Green's function out of equilibrium,  in the following we consider
the adiabatic limit where the current depends only on the
instantaneous equilibrium properties of the dot, i.e. on the
retarded dot Green's function (GF).
This situation is realized in the two following cases. First, let
us consider that only the modulus of the tunneling matrix elements
is varied in time, i.e.
$V_{\beta,j}(t)=|V_{\beta,j}(t)|\exp\{i\Phi_{\beta}\}$. Under the
adiabatic condition, the tunneling rate varies slowly in time, and
the quantum dot can be considered time by time in equilibrium with
the external leads. The effect of quantum pumping is well
described by an adiabatic expansion of the self-energy based on
the average-time approximation as described in
Ref.[\onlinecite{fazio_prl_05}] and using the equilibrium
relations to write the pumped current in terms of the retarded GF
only. Let us consider now the situation in which the modulus of
the tunneling terms is fixed while their phases are modulated in
time, i.e. $V_{\beta,j}(t)=|V_{\beta,j}|\exp\{i\Phi_{\beta}(t)\}$.
This situation is equivalent (by a gauge transformation) to having
a system biased with an ac external signal. In particular, the ac
voltage applied to the leads is proportional to the time
derivative of the tunneling phase
$\partial_t\Phi_{\beta}(t)$\cite{note3}. Since the tunneling terms
are assumed to vary in time with  frequency $\omega$, the ac
signal forcing the quantum dot is proportional to the pumping
frequency $\omega$ and thus can be considered as a small
perturbation under the adiabatic condition. In particular, if one
consider the case in which the tunneling phases vary linearly in
time, $\Phi_{\beta}(t)=\pm V t/2$,  this situation corresponds to
an interacting quantum dot biased by a dc voltage $V$. Following
the work by Meir and Wingreen\cite{meir-wingreen92}, the current
$I$ flowing through the interacting multilevel dot biased by a dc
voltage $V$ can be written as:
\begin{eqnarray}
\label{eq:curr-out-eq}
I=(e/h)\int d\epsilon [f_L(\epsilon)-f_R(\epsilon)]Tr\{G^a\Gamma^R G^r \Gamma^L\mathcal{R}\},
\end{eqnarray}
where $f_{L,R}$ are the Fermi functions, $\Gamma_{L,R}$ are the
dot-leads coupling strengths and $\mathcal{R}=\Sigma_0^{-1}\Sigma$
is the ratio between the fully interacting self-energy and the
noninteracting one, responsible for the deviation from the
Landauer-B\"{u}ttiker formula (see
Ref.[\onlinecite{meir-wingreen92}], Eq.(10)). In the
zero-temperature limit and for a weak bias, $\Sigma=\Sigma_0$ at
the Fermi level and thus Eq. (\ref{eq:curr-out-eq}) can written as
$I\propto V Tr\{G^a\Gamma^R G^r \Gamma^L\} $, which corresponds to
the usual linear response form, even though the Green's function
are interacting ones. This argument is extensively discussed in
Refs.[\onlinecite{meir-wingreen94}] (see Eq.s (37) and (38)) and
[\onlinecite{meir-wingreen92}].

Thus in general we expect that, when both the modulus and the
phase of the pumping parameters are varied in time, apart the
usual dc pumping current a new term arises (that we call of {\it
rectification}) which is proportional to the time derivative of
the tunneling phase: $I_r=\frac{q}{2\pi}Tr\{G^a\Gamma^R G^r
\Gamma^L\}\partial_t(\Phi_L-\Phi_R) $.

Since, as explained above, in adiabatic regime the current is
determined by the instantaneous properties of the dot (retarded
Green's function) and since in the zero temperature limit the
usual linear response formula can be adopted for the calculation
of the current, the pumping and rectification currents through the
interacting quantum dot can be calculated by the scattering matrix
approach as well. In fact, as well known, the retarded Green's
function is related to scattering matrix by the Fisher-Lee
relation. We thus employ the  scattering matrix formalism
developed in Ref.[\onlinecite{wang-wang02}] where the charge
current originated by an adiabatic pump is related to an expansion
of the quantity
$\{S(E,\tau)[f(E+i\partial_{\tau}/2)-f(E)]S^{\dag}(E,\tau)\}_{\alpha\alpha}$
with respect to the time derivative operator $i\partial_{\tau}/2$
(here $S(E,\tau)$ is the scattering matrix and $f(E)$ is the Fermi
function). The first order of this expansion reproduces the famous
Brouwer's formula\cite{brouwer98}. Let us only stress that the
scattering matrix formalism is well defined, not only in the
noninteracting case, but also for the interacting problem (e.g.
see Ref.[\onlinecite{langreth66}]).
The expression of the pumped current $I^\beta(E,\tau)$ in terms of
the time-dependent scattering matrix\cite{brouwer98,wang-wang02}
is:
\begin{eqnarray}
\label{eq:curr} &&
I^{\beta}(E,\tau)=\frac{q }{2\pi}Im\Bigl\{\sum_{\alpha}S_{\beta\alpha}^{\ast}(E,\tau)\partial_{\tau}S_{\beta\alpha}(E,\tau)\Bigl\}, \nonumber \\
&& I^{\beta}(\tau)=\int \frac{d E}{2\pi} (-f')I^{\beta}(E,\tau),
\end{eqnarray}
where $f(E)$ is the Fermi function and $S(E,\tau)$ is
the instantaneous $S$-matrix of the QD.
It is given by the Wigner transform $S(E,\tau)=\int dt e^{iE
t}S(\tau+t/2,\tau-t/2)$, where
\begin{equation}
S_{\alpha\beta}(t,t')=\delta_{\alpha\beta}\delta(t-t')-Tr\{\mathcal{M}^{\alpha\beta}(t,t')\mathbf{G}^r(t,t')\}.
\end{equation}
Here $\textbf{G}^r$ is the full retarded QD Green's function,
$G^r_{n,m}(t,t')=-i\theta(t-t')\langle\{d_n(t),d^\dagger_m(t')\}\rangle$
and $[\mathcal{M}^{\alpha\beta}(t,t')]_{mn}=2\pi i\rho
V^{\ast}_{\alpha,m}(t)V_{\beta,n}(t')$. In the limit of the
pumping frequency $\omega\ll \Gamma$, i.e. under the adiabatic
condition, the scattering matrix $S(E,\tau)$ is expressed by the
instantaneous Green's function of the dot as\cite{note1}:
\begin{equation}
\label{eq:inst} S_{\alpha\beta}(E,\tau)=\delta_{\alpha\beta}-2\pi
i\rho\sum_{n,m}V^{\ast}_{\alpha,m}(\tau)G^r_{nm}(E,\tau)V_{\beta,n}(\tau).
\end{equation}

When substituting (\ref{eq:inst}) into (\ref{eq:curr}) to compute
the current we need the time-derivative of the QD Green's function
which satisfies the relation:
\begin{equation}
\label{eq:der_gf}
\partial_{\tau}\mathbf{G}^r(E,\tau)=\mathbf{G}^r(E,\tau)\mathbf{\dot{\Sigma}}^r(E,\tau)\mathbf{G}^r(E,\tau),
\end{equation}
where the dot symbol indicates a time-derivative and the matrix
notation for the dot Green's function has been used. The final
expression obtained for the $I^\beta(E,\tau)$ for a multi-level
quantum dot is\cite{note2}:
\begin{eqnarray}
\label{eq:main}
 I^{\beta}(E,\tau)&=& \frac{q}{2\pi}[Im\Bigl\{-i
Tr\{\mathbf{\dot{\Gamma}}^{\beta}\mathbf{G}^r\}\Bigl\}
+Im\Bigl\{i(\dot{\Phi}_{\bar{\beta}}-\dot{\Phi}_{\beta})\times\nonumber\\
&\times& Tr\{\mathbf{\Gamma}^{\beta}
\mathbf{G}^a\mathbf{\Gamma}^{\bar{\beta}}\mathbf{G}^r\}\Bigl\}+\Delta^{\beta}(E,\tau)],
\end{eqnarray}
where
\begin{eqnarray}
\Delta^{\beta}(E,\tau)&=&Im\Bigl\{-iTr\{\mathbf{\Gamma}^\beta\mathbf{G}^a\dot{\mathbf{\Sigma}}^r\mathbf{G}^r\}\Bigl\}
\nonumber \\
&+& Im\Bigl\{Tr\{\mathbf{\dot{\Gamma}}^{\beta}\mathbf{G}^r\}Tr\{\mathbf{\Gamma}^{\beta}\mathbf{G}^a\}\Bigl\}\nonumber\\
&+&
\sum_{s,m,n,p}Im\{\Gamma^{\beta}_{sm}G^a_{mn}\Gamma^{\bar{\beta}}_{np}G_{ps}^r\times\nonumber\\
&\times& \partial_{\tau}\ln(|V_{\beta,s}||V_{\bar{\beta},p}|)\}.
\end{eqnarray}
This expression has been obtained by considering explicitly the
time dependence of the modulus and phase of the tunnel matrix
elements, and consequently of the leads-dot coupling function
$\Gamma(t)$. The symbol $\bar{\beta}$ stands for the L,R lead in
correspondence of $\beta$=R,L. The total dc current through the
lead $\beta$ is given by:
\begin{equation}
\label{eq:tot_curr} I^\beta =\frac{\omega}{2\pi}\int
dE\int_0^{2\pi/\omega} d\tau I^{\beta}(E,\tau)[-\partial_E f(E)].
\end{equation}
The expression (\ref{eq:main}) represents our main result. It is
valid for a multi-level QD and for any interaction strength in the
zero temperature limit under the adiabatic condition. The first
term in Eq.(\ref{eq:main}) represents the pumping current, while
the second one, proportional to the time-derivative of the
tunneling rate phase, is the effective rectification term we have
discussed above.
It can also be written as $\mathcal{G}(\tau)V_{eff}(\tau)$, where
$\mathcal{G}(\tau)\propto Tr\{\mathbf{\Gamma}^\beta
\mathbf{G}^a\mathbf{\Gamma}^{\bar{\beta}}\mathbf{G}^r\}$ is the
conductance of the structure, while $V_{eff}(\tau)\propto
(\dot{\Phi}_{\bar{\beta}}-\dot{\Phi}_{\beta})$.
The last term in (\ref{eq:main}) contains information on the time derivative of the retarded self-energy and is zero for a single-level
quantum dot within the wide band limit.

\section{Total current formula for a single level QD}
\label{sec:single-level}

Up to now we have developed a theory of the pumped current valid
in the case of a multi-level QD. We now specialize Eq.
(\ref{eq:main}) to the case of a single level QD. Eliminating the
trace in (\ref{eq:main}) and considering the remaining quantities
as c-numbers,  the expression for the current simplifies to:
\begin{eqnarray}
\label{eq:main-1-level}
&&I^{\beta}(E,\tau)= -\frac{q}{2\pi}[Re\{\dot{\Gamma}^{\beta}G^r\}+\Gamma^{\beta}|G^r|^2\partial_{\tau}Re\{\Sigma^r\}+\nonumber\\
&&+(\dot{\Phi}_{\beta}-\dot{\Phi}_{\bar{\beta}})\Gamma^{\beta}\Gamma^{\bar{\beta}}|G^r|^2].
\end{eqnarray}
When the time-derivative of the tunneling phase is neglected the
above formula is equivalent to the pumped current calculated by
the self-energy adiabatic expansion\cite{fazio_prl_05}.

 In the following we consider the case of a single level QD both
in the strongly interacting and non-interacting case and describe
the behavior of the charge $Q$ (in unit of the electron charge
$q$) pumped per cycle in the zero-temperature limit.
 The Fermi energy $\mu$ is set to zero as
reference energy level, while the static linewidth
$\Gamma_0^{L/R}$ is assumed as energy unit (typical value for
$\Gamma_0^{\alpha}$ is $10 \mu eV$).
\\
In the noninteracting case, i.e. when the QD Green's function
becomes a scalar, the expression for the instantaneous pumping
current is explicitly given by:
\begin{eqnarray}
\label{eq:single-level}
I^\beta(E,\tau)&=&\frac{q}{2\pi}\Bigl[-\frac{(E-\varepsilon_0)\dot{\Gamma}^\beta}{(E-\varepsilon_0)^2+(\Gamma/2)^2}
\nonumber \\
&+&\frac{\Gamma^{\beta}\Gamma^{\bar{\beta}}(\dot{\Phi}_{\bar{\beta}}-\dot{\Phi}_{\beta})}{(E-\varepsilon_0)^2+(\Gamma/2)^2}\Bigl].
\end{eqnarray}
When $\dot{\Phi}_{\bar{\beta}}-\dot{\Phi}_{\beta}=0$,  the charge pumped is
zero when the level is resonant ($\varepsilon_0=0$).

\begin{figure}[htbp]
\centering
\includegraphics[scale=0.42]{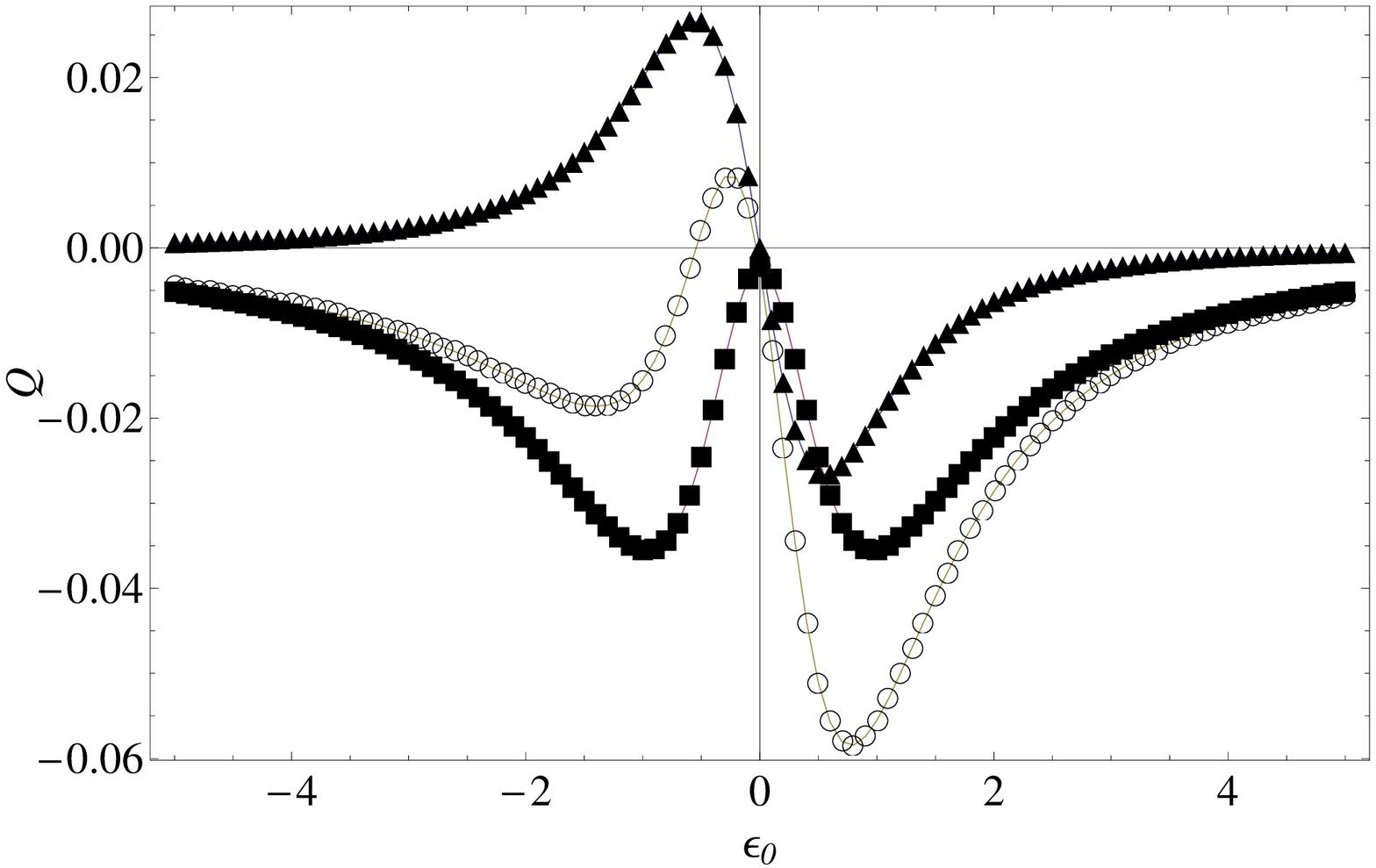}\\
\centering
\includegraphics[scale=0.42]{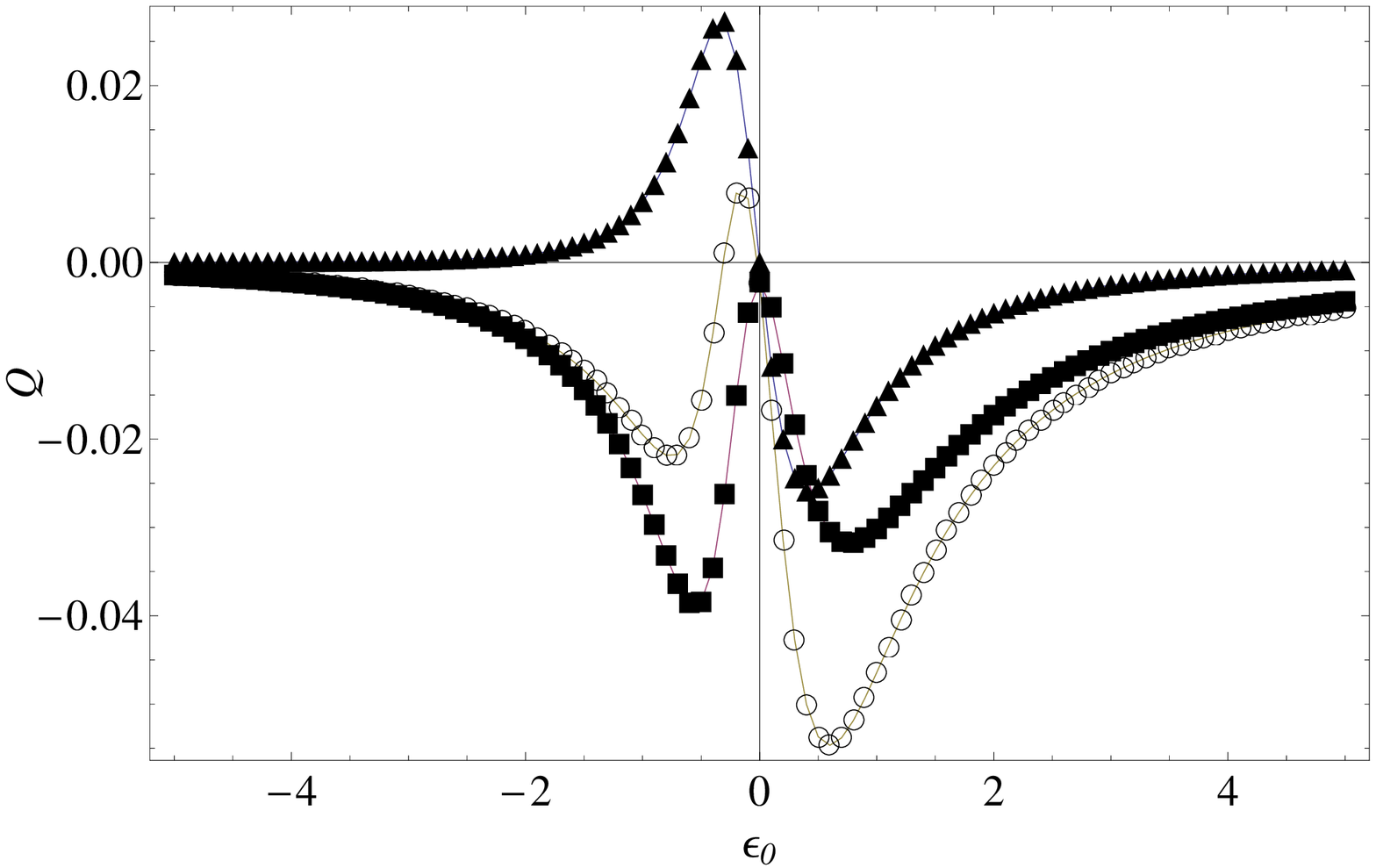}\\
\caption{Charge $Q$ pumped per pumping cycle as a function of the
dot level $\varepsilon_0$. The pumping (triangle) and the
rectification (box) contribution to the total charge (empty
circle) are shown for the non interacting case ($U=0$) in the
upper panel and  for strongly interacting dot ($U\rightarrow\infty$)
in the lower panel. Both figures are computed for the following
choice of parameters: $\mu=0$, $\varphi=\pi/2$,
$\Gamma^{R}_{\omega}=0.2$, $\Gamma^{L}_{\omega}=0.4$,
$\Gamma_0^{L}=\Gamma_0^{R}=1$, $\Phi^{\omega}_{L}=0.3$,
$\Phi^{\omega}_{R}=0.2$. The pumping cycle is determined by:
$\Gamma^{\alpha}(t)=\Gamma_0^{\alpha}+\Gamma^{\alpha}_{\omega}\sin(\omega
t+\varphi_{\alpha})$,
$\Phi_{\alpha}(t)=\Phi^{\omega}_{\alpha}\sin(\omega
t+\varphi_{\alpha})$, with $\alpha=L,R$ and $\varphi_L=0$,
$\varphi_R=\varphi$} \label{fig:fig1}
\end{figure}

\begin{figure}[htbp]
\centering
\includegraphics[scale=0.42]{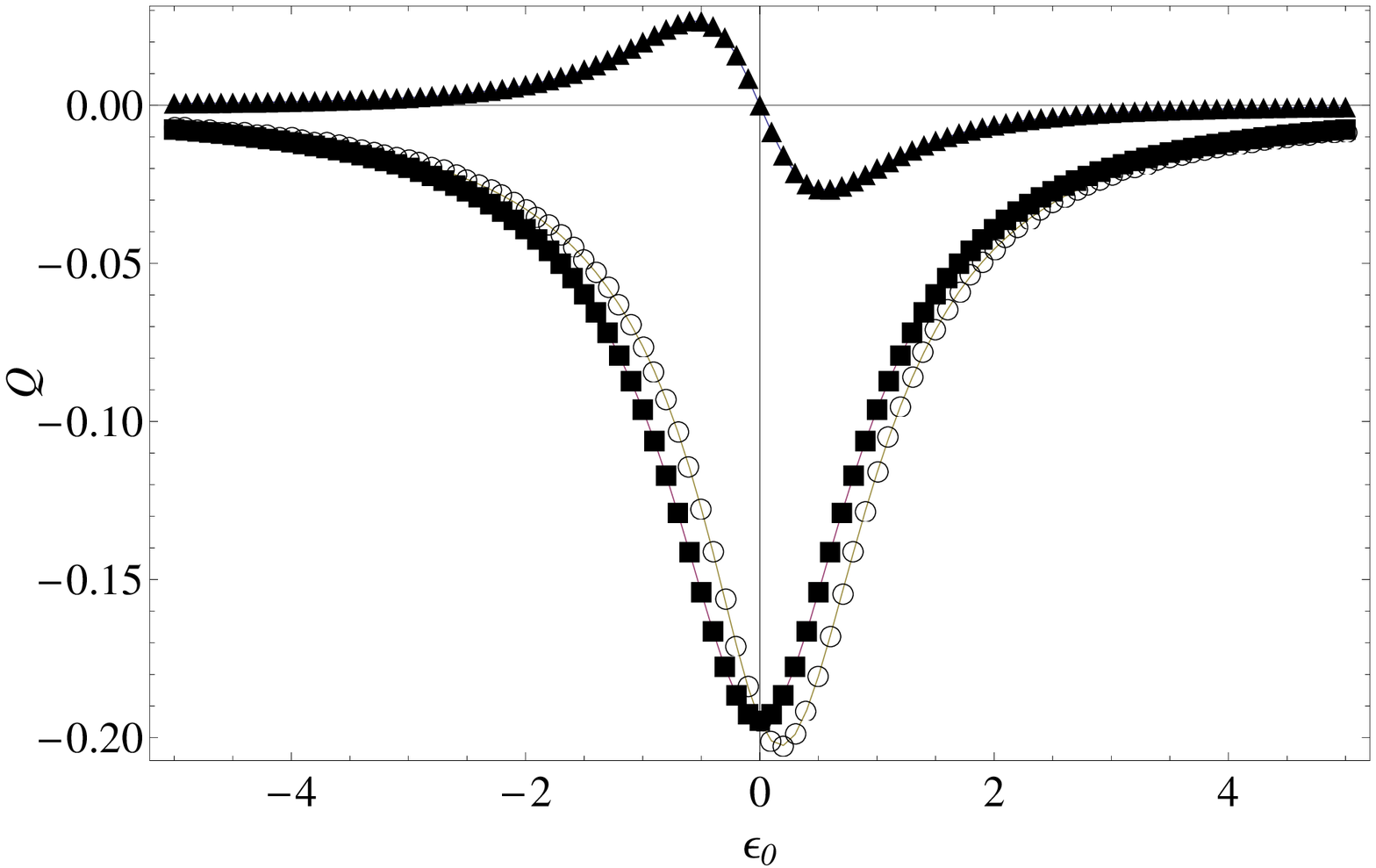}\\
\centering
\includegraphics[scale=0.42]{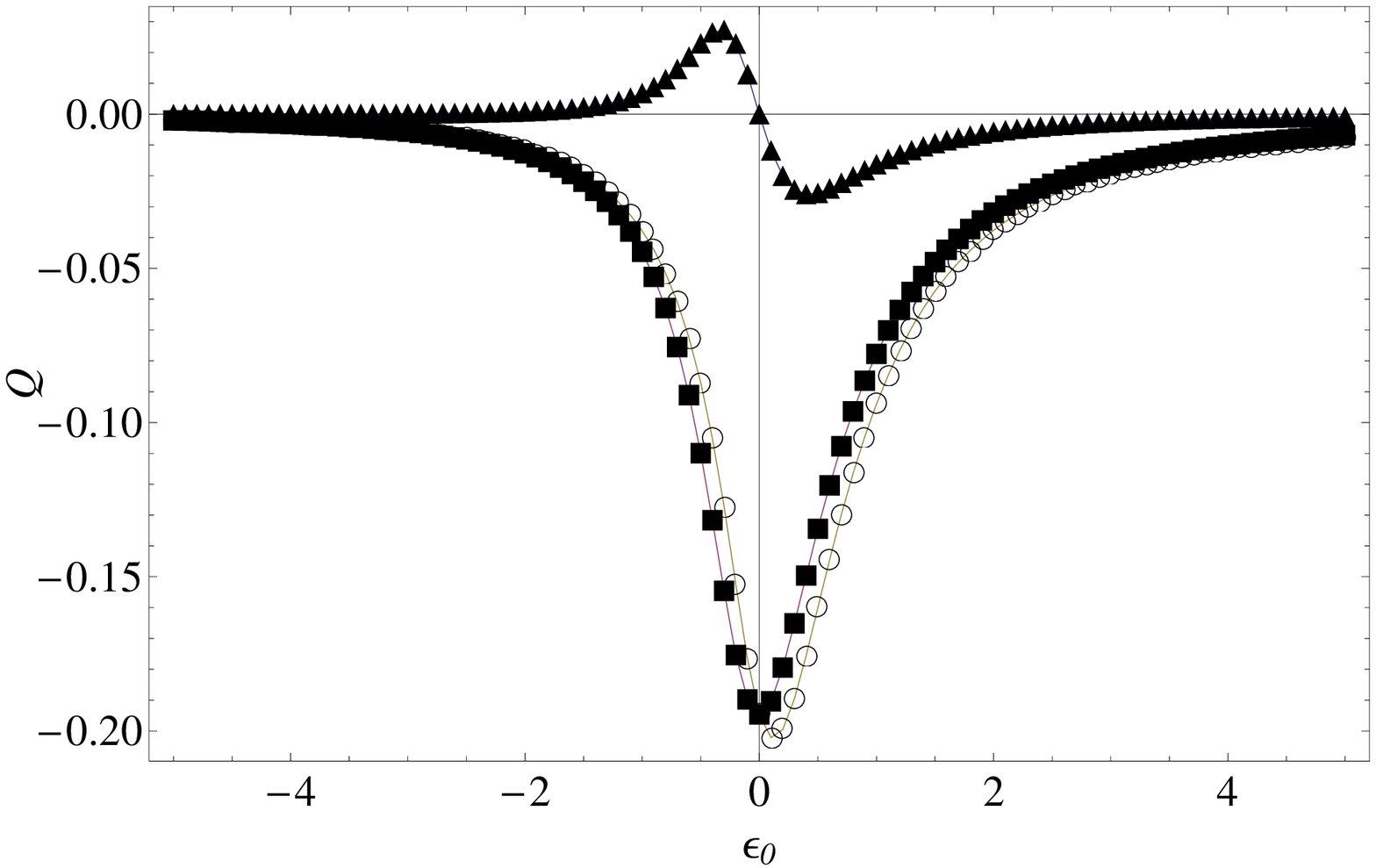}\\
\caption{Charge $Q$ pumped per pumping cycle as a function of the
dot level $\varepsilon_0$ in the non interacting case (upper
panel) and strongly interacting case (lower panel). The pumping
(triangle) and the rectification (box) contribution to the total
charge (empty circle) are shown in the zero temperature-limit by
setting the remaining parameters as follows: $\mu=0$,
$\varphi=\pi/2$, $\Gamma^{R}_{\omega}=0.2$,
$\Gamma^{L}_{\omega}=0.4$, $\Gamma_0^{L}=\Gamma_0^{R}=1$,
$\Phi^{\omega}_{L}=0.3$, $\Phi^{\omega}_{R}=0.2$. Differently from
Fig.\ref{fig:fig1}, the pumping cycle is determined by:
$\Gamma^{\alpha}(t)=\Gamma_0^{\alpha}+\Gamma^{\alpha}_{\omega}\sin(\omega
t+\varphi_{\alpha})$, $\Phi_{\alpha}(t)=\Phi^{\omega}_{\alpha}t$,
with $\alpha=L,R$ and $\varphi_L=0$, $\varphi_R=\varphi$.} \label{fig:fig2}
\end{figure}

\begin{figure}[htbp]
\centering
\includegraphics[scale=0.52]{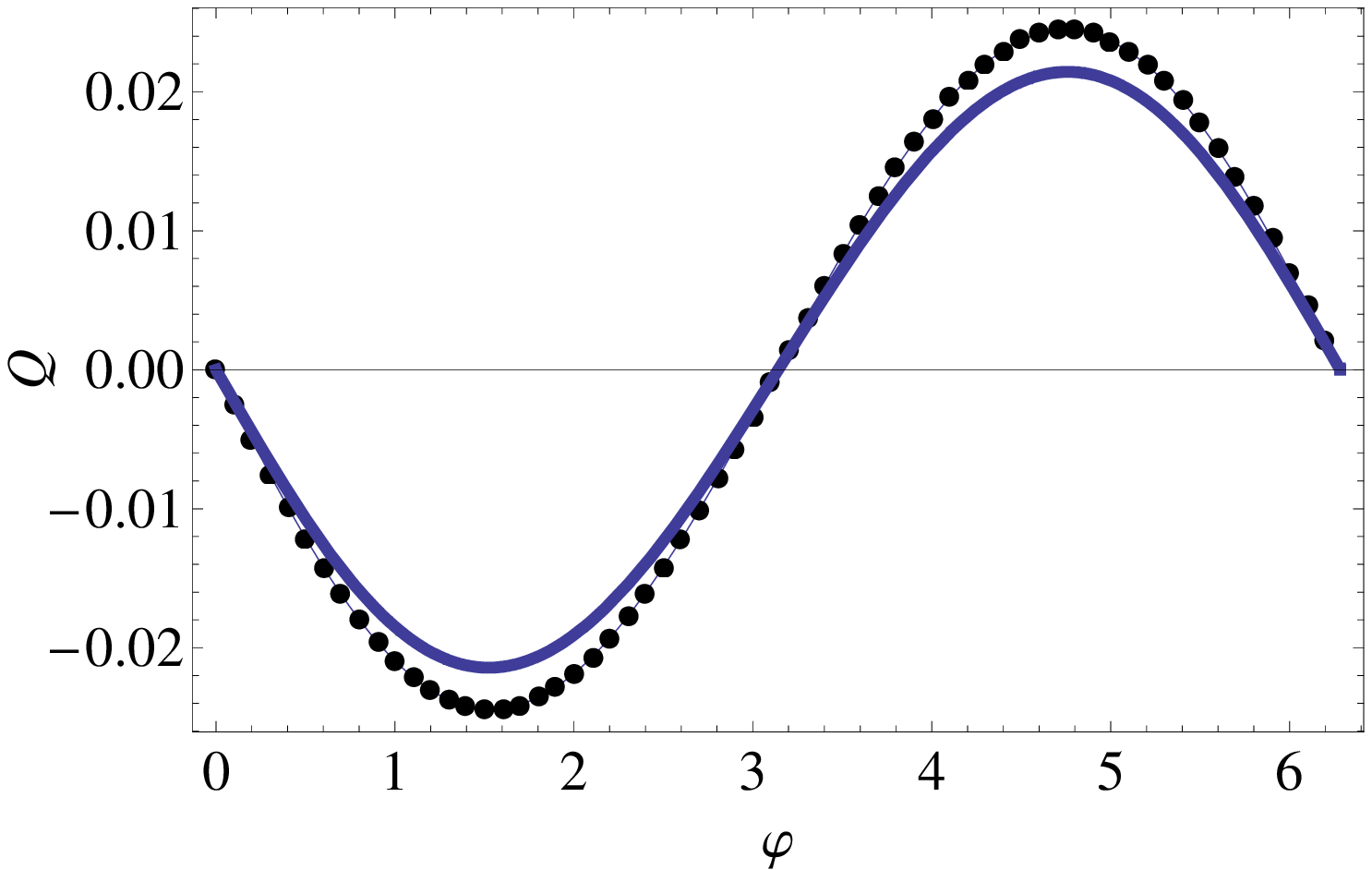}\\
\includegraphics[scale=0.52]{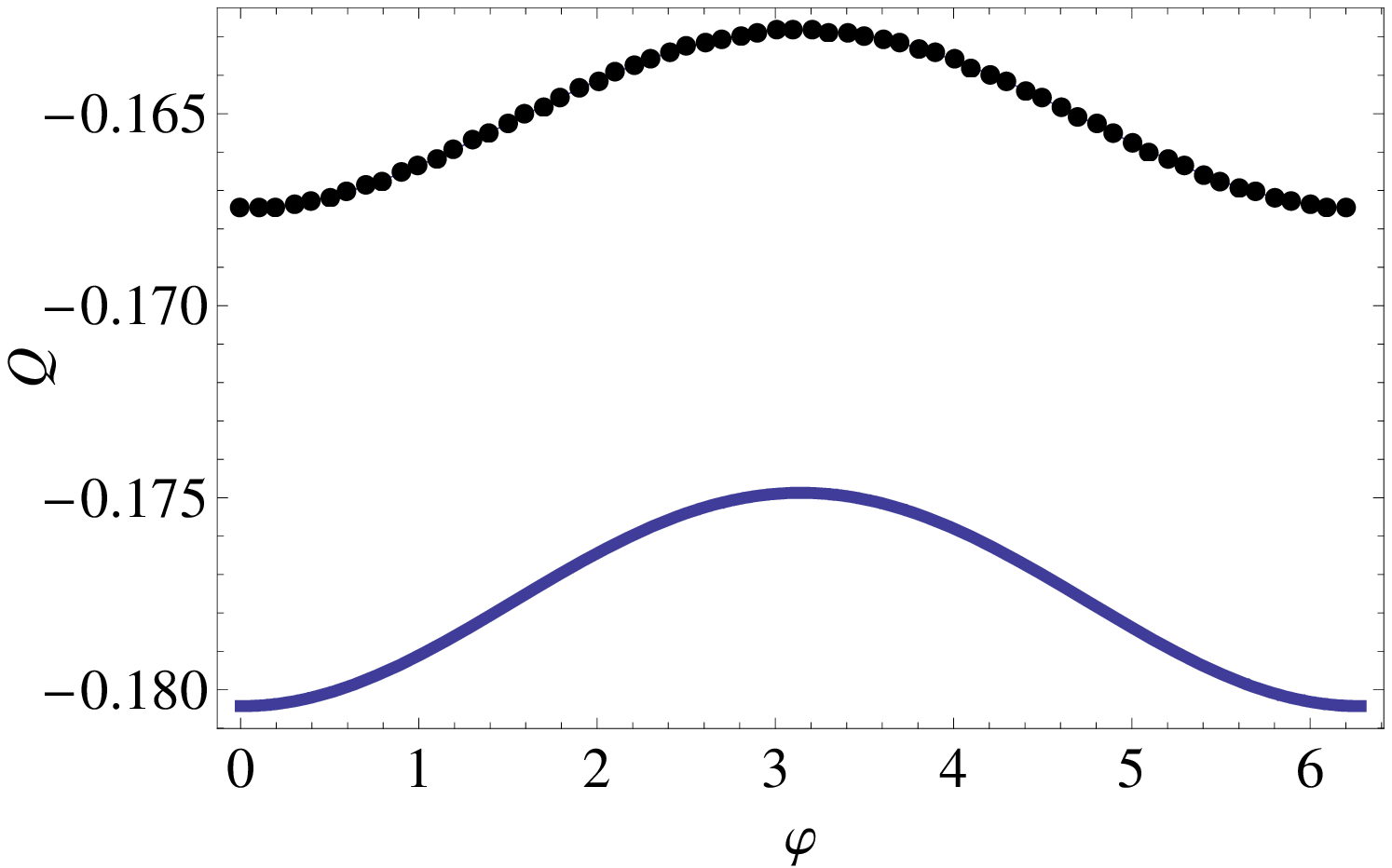}\\
\caption{The behavior of the pumping current (upper panel) and of
the rectification current (lower panel) with respect to $\varphi$
is reported for $\varepsilon_0=0.3$ and taking the remaining
parameters as in Fig.\ref{fig:fig2}. Notice that the pumping term
is proportional to $\sin(\varphi)$, while the rectification one is
proportional to $\cos(\varphi)$. Both in the upper and lower
panel, the full line represents the non interacting result, while
the interacting case is indicated by full circles ($\bullet$).}
\label{fig:fig3}
\end{figure}

In the case of a strongly interacting quantum dot, i.e. in the
infinite-$U$ limit, we take the expression of the QD Green's
function as in Ref.[\onlinecite{souza07}]. The current is:

\begin{eqnarray}
\label{eq:single-level-interact}
I_{\sigma}^\beta(E,\tau)&=&\frac{q}{2\pi}\Bigl[-\frac{(E-\varepsilon_0)(1-n_{\bar{\sigma}})\dot{\Gamma}^\beta}{(E-\varepsilon_0)^2+(\Gamma/2)^2(1-n_{\bar{\sigma}})^2}
\nonumber \\
&+&\frac{\Gamma^{\beta}\Gamma^{\bar{\beta}}(\dot{\Phi}_{\bar{\beta}}-\dot{\Phi}_{\beta})(1-n_{\bar{\sigma}})^2}{(E-\varepsilon_0)^2+(\Gamma/2)^2(1-n_{\bar{\sigma}})^2}\Bigl],
\end{eqnarray}
where the occupation number $n_{\sigma}$ on the dot has to be
determined self-consistently by the relation $n_{\sigma}=(2\pi
i)^{-1}\int dE G_{\sigma\sigma}^{<}(E)$, where
$G_{\sigma\sigma}^{<}$ is the QD lesser Green's function.\\
In order to show the effects of the time-dependent tunneling phase
we report below the numerical results of the charge $Q$ pumped per
cycle. The pumping cycle is determined by the periodic time
variation of the leads-dot coupling strength, where
$\Gamma^{\alpha}(t)=\Gamma_0^{\alpha}+\Gamma^{\alpha}_{\omega}\sin(\omega
t+\varphi_{\alpha})$, while for the tunneling phase two cases can
be considered. Either it varies harmonically
$\Phi_{\alpha}(t)=\Phi^{\omega}_{\alpha}\sin(\omega
t+\varphi_{\alpha})$ with the same frequency of the two external
gate voltages (this case is shown in Fig.\ref{fig:fig1}) or it
varies linearly in time
$\Phi_{\alpha}(t)=\Phi^{\omega}_{\alpha}t$, e.g. when a
parasitic gate voltage is present, (this case is shown
in Fig.\ref{fig:fig2}). The quantity $\varphi_\alpha$ is the pumping phase that we take different from zero between L and R lead.\\
In Fig.\ref{fig:fig1} we plot $Q$  as a function of the energy
level $\varepsilon_0$ by fixing the other parameters as: $\mu=0$,
$\varphi=\pi/2$, $\Gamma^{R}_{\omega}=0.2$,
$\Gamma^{L}_{\omega}=0.4$, $\Gamma_0^{L}=\Gamma_0^{R}=1$,
$\Phi^{\omega}_{L}=0.3$, $\Phi^{\omega}_{R}=0.2$. In particular,
in the upper panel, the charge induced by the pumping (triangle)
and the charge due to the rectification term (box) is shown for a
non-interacting dot ($U=0$). The total charge (empty circle) is
significantly modified by the presence of the rectification term
which is a non-vanishing quantity at the Fermi energy
($\varepsilon_0=0$). The lower panel in Fig.\ref{fig:fig1} shows
the behavior of the rectification current as a function of
$\varepsilon_0$ in the strongly interacting limit ($U\rightarrow
\infty$) and by choosing the remaining parameters as in the upper
panel. While the general aspect of the total charge (empty circle)
is only marginally modified by the strong correlations in the
specified region of parameters, the rectified charge (box) shows a
pronounced asymmetric behavior with respect to the level of the
dot.
In Fig.\ref{fig:fig2} we focus on the case of time-linear
variation of the phase $\Phi_{\alpha}(t)=\Phi_{\alpha}^{\omega}t$,
and take the parameters as in Fig.\ref{fig:fig1}. In the upper
panel, the pumped charge $Q$ is plotted as a function of the dot
level $\varepsilon_0$ in the non-interacting case ($U=0$).
Contrary to the previous case, the rectification contribution
(box) is dominant over the one induced by the pumping mechanism
(triangle) and thus the total charge (empty circle) is mainly
affected by a resonant-like behavior. In the lower panel, the
results for $U\rightarrow\infty$ are shown. Apart from a
renormalization of the linewidth of the resonance induced by the
factor $(1-n_{\bar{\sigma}})$ in the numerator of Eq.
(\ref{eq:single-level-interact}), a behavior similar to the one of
the non-interacting system is found. A slave boson treatment with
the inclusion of a renormalization of the dot energy level, could
in principle modify this picture. Let us note that when the
tunneling phase varies harmonically both the pumping current and
the rectification current follow the same $\sin(\varphi)$ behavior
w.r.t. the pumping phase $\varphi$.
In Fig.\ref{fig:fig3} we show the behavior of the charge induced
by the pumping term (upper panel) and by the rectification (lower
panel) with respect to the pumping phase $\varphi$ and by fixing
the dot level to $\varepsilon_0=0.3$ and the remaining parameters
as done in Fig.\ref{fig:fig2}. The full line in both panels
represents the result for the $U=0$ case, while the full circles
($\bullet$) represent
 the curves computed for the infinite-$U$ case. Let us note that while the pumping term follows the
conventional $\sin(\varphi)$-behavior as in Brouwer theory, the
rectification contribution takes the form
$V_{eff}[\mathcal{A}+\mathcal{B}\cos(\varphi)]$, where the
coefficients $\mathcal{A}$ and $\mathcal{B}$ for the $U=0$ case
are explicitly given by:
\begin{eqnarray}
&&\mathcal{A}=
\frac{\Gamma_0^L\Gamma_0^R}{\xi+(\Gamma_0/2)^2}\\\nonumber &&
\mathcal{B}=-\frac{\Gamma^L_{\omega}\Gamma^R_{\omega}[\Gamma_0^2(\Gamma_0^L-\Gamma_0^R)^2+2\Gamma_0^L\Gamma_0^R(4\xi-\Gamma_0^2)-16\xi^2]}{32[\xi+(\Gamma_0/2)^2]^3},
\end{eqnarray}
with $\xi=(\mu-\varepsilon_0)^2$. The different symmetry of the
pumping and rectification current has already been reported in
experimental works in quantum dots\cite{watson-pump} and some
theoretical explanations have been proposed\cite{theor_nostro}.
Furthermore, the analysis of the coefficient $\mathcal{B}$ shows
that the charge transferred by the rectification effect can be
significantly increased by coupling the dot region to the leads in
an asymmetric way ($\Gamma_0^L\neq \Gamma_0^R$), e.g. by using
tunnel barriers with very different transparencies. The above
results remain almost unchanged in the strongly interacting case
($U\rightarrow \infty$). \\
In the limiting case in which the phase difference between the
tunneling barriers is kept zero, i.e. $\varphi_L=\varphi_R=0$, the
pumping term is exactly zero and the rectification contribution
acts as a quantum ratchet\cite{exp_ratch}.

\section{Conclusions}
\label{sec:conclusions}
Within the Green's function and scattering matrix approach we have
analyzed the quantum pumping current through an interacting
quantum dot when both the modulus and the phase of the model
time-dependent parameters, in our case the leads-dot tunneling
rate, is adiabatically varied. In this way it has been possible to
derive an expression for the pumped current containing an
effective rectification term due to the time-dependent phase. Such
contribution can be written in a Landauer-Buttiker-like form, even
though for the interacting system, when the zero temperature limit
and the adiabatic conditions are met.
The numerical analysis also show that when the tunneling phase
varies linearly in time the rectification term is even with
respect to the pumping phase $\varphi$, i.e. of the form
$a+b\cos(\varphi)$, in contrast to the usual pumping term which is
odd. The mentioned contribution
could be related to the experimentally observed rectification effects in quantum dots \cite{watson-pump}.\\
In particular, we have been considering an open system, but
concerning  closed systems (e.g. annular devices with a quantum
dot), the tunneling phase contribution to the pumped charge could
find a natural interpretation in a complex phase of geometric
nature\cite{whitney_05}. This phase would be a Berry phase
\cite{berry-closed-loop}. Thus the detection of a rectification
current in addition to the pumping one could be an indirect probe
of a Berry phase.\\
The proposed analysis could be easily generalized up to the second
order in the pumping frequency $\omega$ allowing to describe
features involved in the moderate non-adiabatic limit.

\section{Acknowledgements}
We thank Dr. Adele Naddeo for useful suggestions. We regret to
acknowledge the passing away of Prof. Maria Marinaro with whom we
have shared enlightening discussions during the completion of this
work .

\bibliographystyle{prsty}

\begin{thebibliography}{99}
\bibitem{note1} In formula (\ref{eq:inst}) the sum on the discretized
dot levels has been explicitly written.
\bibitem{note2} In deriving the expression for the instantaneous
current we have been using the relation $
\mathbf{G}^r-\mathbf{G}^a=-i\mathbf{G}^a\mathbf{\Gamma}
\mathbf{G}^r$ for the istantaneous Green's function
$G^r_{n,m}(E,\tau)$ of the QD, while we defined
$\Gamma^\alpha_{nm}(\tau,\tau)=2\pi\rho V_{\alpha,m}(\tau)
V^{\ast}_{\alpha,n}(\tau)$.
\bibitem{note3} When an external d.c. or a.c. bias voltage is applied to the system, a gauge
transformation can be perfomed to eliminate the bias and the
tunneling term acquires a time-dependent phase proportional to
the bias. In our case the time-dependent phase of the tunneling
comes from the representation of a complex quantity.
\end{thebibliography}

\end{document}